\documentclass[aps,twocolumn,preprintnumbers,nofootinbib,superscriptaddress]{revtex4-1}
\usepackage{amsmath} \usepackage{graphicx} \usepackage{amsfonts}
\usepackage{array} \usepackage{amsthm} \usepackage{bm}
\usepackage{palatino} \usepackage{mathpazo} 
\usepackage{supertabular}

\usepackage[breaklinks]{hyperref}
\usepackage{color}

\newcommand{\be}{\begin{equation}}
\newcommand{\ee}{\end{equation}}
\newcommand{\ba}{\begin{eqnarray}}
\newcommand{\ea}{\end{eqnarray}}
\newcommand{\bal}{\begin{align}}
\newcommand{\eal}{\end{align}}

\newcommand{\bw}{\begin{widetext}}
	\newcommand{\ew}{\end{widetext}}

\begin{document}
	\title{{On {\textquoteleft Rotating charged AdS solutions in quadratic $f(T)$ gravity\textquoteright}: New rotating solutions}}

	\author{Mustapha Azreg-A\"{\i}nou}
	\affiliation{Ba\c{s}kent University, Engineering Faculty, Ba\u{g}l{\i}ca Campus, 06790-Ankara, Turkey}


	\begin{abstract}
		We show that there are two or more procedures to generalize the known four-dimensional transformation, aiming to generate cylindrically rotating charged exact solutions, to higher dimensional spacetimes . In the one procedure, presented in Eur. Phys. J. C (2019) \textbf{79}:668, one uses a non-trivial, non-diagonal, Minkowskian metric $\bar{\eta}_{ij}$ to derive complicated rotating solutions. In the other procedure, discussed in this work, one selects a diagonal Minkowskian metric ${\eta}_{ij}$ to derive much simpler and appealing rotating solutions. We also show that if ($g_{\mu\nu},\,\eta_{ij}$) is a rotating solution then ($\bar{g}_{\mu\nu},\,\bar{\eta}_{ij}$) is a rotating solution too with similar geometrical properties, provided $\bar{\eta}_{ij}$ and ${\eta}_{ij}$ are related by a symmetric matrix $R$: $\bar{\eta}_{ij}={\eta}_{ik}R_{kj}$.
	\end{abstract}
	
	
	\maketitle
	
\section*{Preliminaries}
In this work we will use the notation of~\cite{3} with a slight difference. Instead of taking $f(T)=T+\alpha T^2$ with $\alpha<0$ we will take $f(T)=T-\alpha T^2$ with $\alpha>0$. 
	
Another different choice, which will be made clearer later, is the signature of the $N$-dimensional Minkowski spacetime: $(+,-,-,-,\cdots)$. Most of the other notations will be almost similar to that of~\cite{3}.
	
As a first comment we state that there are some sign mistakes in the definition of $K_{\alpha\mu\nu}$ of~\cite{3}. We use the following definitions\footnote{$S^{\alpha\mu\nu}$ may be given in a more compact form as:$$S^{\alpha\mu\nu}=\frac{1}{4}(T^{\nu\mu\alpha}+T^{\alpha\mu\alpha}-T^{\mu\nu\alpha})-\frac{1}{2}g^{\alpha\nu}{T^{\sigma\mu}}_{\sigma}+\frac{1}{2}g^{\alpha\mu}{T^{\sigma\nu}}_{\sigma}.$$}:
\begin{align}\label{o1}
	&{T^\alpha}_{\mu \nu}={e_b}^\alpha
	(\partial_\mu{e^b}_\nu-\partial_\nu{e^b}_\mu),\nonumber\\
	&K_{\alpha\mu\nu}=\frac{1}{2}~(T_{\mu\alpha\nu}+T_{\nu\alpha\mu}-T_{\alpha\mu\nu}),\nonumber\\
	&S^{\alpha\mu\nu}=\frac{1}{2}~(K^{\mu\nu\alpha}-g^{\alpha\nu}{T^{\sigma\mu}}_{\sigma}+g^{\alpha\mu}{T^{\sigma\nu}}_{\sigma}),\nonumber\\
	&T=T_{\alpha\mu\nu}S^{\alpha\mu\nu}.
\end{align}
It is obvious from these definitions that the global sign of $T$ would depend on the signature of the metric. For a static metric with signature $(+,-,-,-,\cdots)$
	\begin{equation}\label{m}
	{\rm d}s^2=A(r){\rm d}t^2-\frac{1}{B(r)}{\rm d}r^2-r^2\Big(\sum_{i=1}^{n}{\rm d}\phi^2_i+\sum_{i=1}^{N-n-2}{{\rm d}z^2_i\over l^2}\Big),
	\end{equation}
	where $n$ is the number of angular coordinates, $N$ is the dimension of spacetime and $l$ is related to the cosmological constant by 
	\begin{equation}\label{L}
	\Lambda=-\frac{(N-1)(N-2)}{2l^2}<0.
	\end{equation}
	We obtain
	\begin{equation}\label{T}
	T=+\frac{(N-2)A'B}{rA}+\frac{(N-2)(N-3)B}{r^2}.
	\end{equation}
	Had we reversed the signature of the metric we would obtain the same expression with the two {\textquoteleft $+$\textquoteright} sings changed to {\textquoteleft $-$\textquoteright} sings. A second comment is also in order: The expression of $T$ given in~\cite{3} has an extra factor 2 in the term including $A'$.
	
	{A final comment: The last term in Eq.~(14) of~\cite{3} should have the opposite global sign. Using our metric-signature choice, Eq.~(14) of~\cite{3} takes the form
		\begin{multline}\label{m14}
		{\rm d}s^2=A(r)\Big(\Xi{\rm d}t-\sum_{i=1}^{n}\omega_i{\rm d}\phi_i\Big)^2-\frac{{\rm d}r^2}{B(r)}-r^2\sum_{i=1}^{N-n-2}{{\rm d}z^2_i\over l^2}\\
		\hspace{-1mm}-\frac{r^2}{l^4}\sum_{i=1}^{n}(\omega_i{\rm d}t-\Xi l^2{\rm d}\phi_i)^2-\frac{r^2}{l^2}\sum_{i<j}^{n}(\omega_i{\rm d}\phi_j-\omega_j{\rm d}\phi_i)^2 ,
		\end{multline}
		where ($\omega_1,\,\omega_2,\cdots,\,\omega_n$) are the rotation $n$ parameters, ($\phi_1,\,\phi_2,\cdots,\,\phi_n$) are the $n$ angular coordinates and $\Xi=\sqrt{1+\Sigma_{i=1}^n(\omega_i^2/l^2)}$. Note that the last term, $-(r^2/l^2)\sum_{i<j}^{n}(\omega_i{\rm d}\phi_j-\omega_j{\rm d}\phi_i)^2$, vanishes identically if the spacetime has only one  angular coordinate.}
	
	The field equations of Maxwell-$f(T)$ gravity are given in Eq.~(3) of~\cite{3}, which we rewrite here for convenience
	\begin{align}\label{o2}
	& I_\mu{^\nu}\equiv {S_\mu}^{\rho \nu} \partial_{\rho} T
	f_{TT}+\Big[e^{-1}{e^a}_\mu\partial_\rho\Big(e{e_a}^\alpha
	{S_\alpha}^{\rho \nu}\Big)-{T^\alpha}_{\lambda \mu}{S_\alpha}^{\nu \lambda}\Big]f_T\nonumber\\
	&-\frac{\delta^\nu_\mu}{4}\Big(f+\frac{(N-1)(N-2)}{l^2}\Big) +{\kappa \over 2} T_{\text{(em)}\,\mu}{}^\nu=0,\nonumber\\
	& \partial_\nu \Big(\sqrt{|g|} F^{\mu \nu} \Big)=0,
	\end{align}
	where $e\equiv \sqrt{|g|}$ and $T_{\text{(em)}\,\mu}{}^\nu=F_{\mu \alpha}F^{\nu \alpha}-\frac{1}{4} \delta_\mu{}^\nu F_{\alpha \beta}F^{\alpha \beta}$, with $F_{\alpha \beta}=\partial_\alpha A_\beta-\partial_\beta A_\alpha$, is the energy-momentum tensor of the electromagnetic field. Here the ratio $(N-1)(N-2)/l^2$ is proportional to the cosmological constant $\Lambda$~\eqref{L}. It is obvious from the shape of Eqs.~\eqref{o2} that we are dealing with a spin-zero (pure tetrad) $f(T)$ gravity. The general field equations including spin connection terms are provided in~\cite{nonzero}.
	
	A particular charged static solution to the field equations~\eqref{o2} with $f(T)=T-\alpha T^2$ and $\alpha=-1/(24\Lambda)>0$ has been determined~\cite{Awad} and is given by Eqs.~(8) of~\cite{3}
	\begin{align}\label{sol}
	&A(r)=\frac{r^2}{6(N-1)(N-2)\alpha}-\frac{m}{r^{N-3}}+\frac{3(N-3)q{}^2}{(N-2)r^{2(N-3)}}\nonumber\\
	&\qquad +\frac{2\sqrt{6\alpha}(N-3)^3q{}^3}{(2N-5)(N-2)r^{3N-8}},
	\nonumber\\
	&B(r)=A(r)\Big[1+\frac{\sqrt{6\alpha}(N-3)q}{r^{N-2}}\Big]^{-2},\nonumber\\
	&\Phi(r)=\frac{q}{r^{N-3}}+\frac{\sqrt{6\alpha}(N-3)^2q{}^2}{(2N-5)r^{2N-5}}, 
	\end{align}
	where we have replaced $\Lambda_{\text{eff}}$ by $1/[6(N-1)(N-2)\alpha]$. Note that since $\alpha>0$ we have $\Lambda_{\text{eff}}>0$.
	
	{\section*{Generating cylindrically rotating charged exact solutions}}
	{Consider the following substitution where $a$ denotes a rotation parameter
		\begin{equation}\label{sub}
		{\rm d}t\to \sqrt{1+\frac{a^2}{l^2}}~{\rm d}t - a {\rm d}\phi ,\quad {\rm d}\phi\to \sqrt{1+\frac{a^2}{l^2}}~{\rm d}\phi - \frac{a}{l^2} {\rm d}t .
		\end{equation}
		There is no claim whatsoever in Refs.~\cite{2,2b} that the substitution~\eqref{sub} is a shortcut or a trick for generating rotating solutions from static ones, however, some authors have applied the substitution~\eqref{sub} as a procedure to generate their supposed-to-be rotating solutions. In this work we present a general comment on the transformation~\eqref{sub} and its generalization to higher dimensions.}
	
	Our starting point is the expression of the tetrad $e^i{}_{\mu}$ in terms of the static metric $(A(r),\,B(r))$, the $n$ rotation parameters denoted by ($\omega_1,\,\omega_2,\cdots,\,\omega_n$) instead of ($a_1,\,a_2,\cdots,\,a_n$), and the constant $\Xi=\sqrt{1+\Sigma_{i=1}^n(\omega_i^2/l^2)}$. The tetrad expression $e^i{}_{\mu}$ is given in Eq.~(12) of~\cite{3}. {However, in order to evaluate $e_i{}^{\mu}$ from $e^i{}_{\mu}$, using the expression $e_i{}^{\mu}=\eta_{i j}g^{\mu \nu}e^j{}_{\nu}$, we need an expression for the Minkowskian metric $\eta_{i j}$. The authors of Ref.~\cite{3} \emph{did not provide any expression for} $\eta_{i j}$ they used in their work. An anonymous referee claimed that it is the non-diagonal form of $\eta_{i j}$, as given in Eq.~(44) of Ref.~\cite{4} and Eq.~(41) of Ref.~\cite{5}, that has been used in~\cite{3} and that it is the only valid form of $\eta_{i j}$ to be used. In this work we will use two different expressions for $\eta_{i j}$ and we shall show that \emph{the statement of the referee does not hold true} by constructing a new cylindrically rotating charged solution using a \emph{diagonal} expression for $\eta_{i j}$.}
	
	For $N=4$ we have checked that the proposed rotating solution in~\cite{3} satisfies the field equations~\eqref{o2} with $\kappa=-2$ {taking a diagonal Minkowskian metric $\eta_{i j}=\text{diag}(1,\,-1,\,-1,\,-1)$.
	
	From now on we restrict ourselves to $N=5$ and consider the cases 1) $n=1$ and 2) $n=2$.\\
	
	\subsection*{Case 1) $N=5,\,n=1$}
	In this case the coordinates are denoted by ($t,\,r,\,\phi,\,z_1,\,z_2$). The tetrad expression~(12) of~\cite{3} reduces to
	\begin{equation}\label{r1}
	({e^i}_\mu)=\left(
	\begin{array}{ccccc}
	\Xi  \sqrt{A(r)} & 0 & -\omega  \sqrt{A(r)} & 0 & 0 \\
	0 & \frac{1}{\sqrt{B(r)}} & 0 & 0 & 0 \\
	\frac{\omega  r}{l^2} & 0 & -\Xi  r & 0 & 0 \\
	0 & 0 & 0 & \frac{r}{l} & 0 \\
	0 & 0 & 0 & 0 & \frac{r}{l}
	\end{array}
	\right).
	\end{equation}
	This is not a proper tetrad as the associated spin connection does not vanish~\cite{nonzero,nonzero2}. To evaluate the associated spin connection we refer to~\cite{nonzero,nonzero2}. Using the terminology of these references, the reference tetrad ${e_{\text{(r)}\mu}^i}$ is, in this case, given by~\eqref{r1} upon setting $m=q=0$ (absence of gravity and matter) and $N=5$. We find that the nonvanishing components of the spin connection $\omega^a{}_{b\mu}$ are [the Latin indexes ($a,\,b$) in $\omega^a{}_{b\mu}$ run from 1$\to$5]: $\omega^1{}_{2t}=\omega^2{}_{1t}=\Xi r/(72\alpha)$, $\omega^1{}_{2\phi}=\omega^2{}_{1\phi}=-\omega r/(72\alpha)$, $\omega^2{}_{3t}=\omega^3{}_{2t}=-\omega r/(6\sqrt{2\alpha}l^2)$, $\omega^2{}_{3\phi}=\omega^3{}_{2\phi}=-\Xi r/(6\sqrt{2\alpha}l^2)$, $\omega^2{}_{4z_1}=\omega^4{}_{2z_1}=\omega^2{}_{5z_2}=\omega^5{}_{2z_2}=-r/(6\sqrt{2\alpha}l)$. This fact results in violation of local Lorentz invariance.
	
	Taking a diagonal Minkowskian metric $\eta_{i j}=\text{diag}(1,\,-1,\,-1,\,-1,\,-1)$, the corresponding metric $g_{\mu \nu} =
	\eta_{ij}e^{i}{_{\mu}}e^{j}{_{\nu}}$ reads
	\begin{multline}\label{r1b}
	{\rm d}s^2=A(r) (\Xi  {\rm d} t-\omega  {\rm d} \phi )^2-\frac{{\rm d} r^2}{B(r)}\\
	-\frac{r^2}{l^4} \Big(\omega  {\rm d} t-\Xi  l^2 {\rm d} \phi \Big)^2-\frac{r^2{\rm d} z_1^2}{l^2}-\frac{r^2{\rm d}
		z_2^2}{l^2},
	\end{multline}
	which is the same as the metric suggest in Eq.~(14) of~\cite{3}; in this case ($N=5,\,n=1$) the last term in Eq.~(14) of~\cite{3} vanishes identically.
	
	Now, we evaluate $T$ upon substituting~\eqref{r1} and~\eqref{r1b} into~\eqref{o1} and the resulting expression is identical to~\eqref{T} taking $N=5$.
	
	On substituting~\eqref{r1}, \eqref{r1b} and~\eqref{T} into the field equations~\eqref{o2} and using the static solution~\eqref{sol} we noticed that all the field equations are satisfied.

	\subsection*{Case 2) $N=5,\,n=2$}
	In this case the coordinates are denoted by ($t,\,r,\,\phi_1,\,\phi_2,\,z$). The tetrad expression~(12) of~\cite{3} reduces to
	\begin{equation}\label{r4}
	({e^i}_\mu)=\left(
	\begin{array}{ccccc}
	\Xi  \sqrt{A(r)} & 0 & -\omega _1 \sqrt{A(r)} & -\omega _2 \sqrt{A(r)} & 0 \\
	0 & \frac{1}{\sqrt{B(r)}} & 0 & 0 & 0 \\
	\frac{\omega _1 r}{l^2} & 0 & -\Xi  r & 0 & 0 \\
	\frac{\omega _2 r}{l^2} & 0 & 0 & -\Xi  r & 0 \\
	0 & 0 & 0 & 0 & \frac{r}{l}
	\end{array}
	\right).
	\end{equation}
	
	{In order to proof that Eq.~(14) of Ref.~[1], which is Eq.~\eqref{m14} of this work (including the global sign correction we made), is a rotating solution one needs an expression for the Minkowskian matrix $\eta_{ij}$ by which one can evaluate $e_i{}^{\mu}$ from $e^i{}_{\mu}$~\eqref{r4}, then evaluate all the tensors needed in the field equations~\eqref{o2}. We divide this case into two sub-cases a) $\eta_{i j}$ diagonal and b) $\eta_{i j}$ non-diagonal.
	
	\subsubsection*{{Case a) $\eta_{i j}$ diagonal}}
	{If $\eta_{i j}=\text{diag}(1,\,-1,\,-1,\,-1,\,-1)$,} the corresponding metric $g_{\mu \nu} =
	\eta_{ij}e^{i}{_{\mu}}e^{j}{_{\nu}}$ takes the form
	\begin{multline}\label{r4b}
	{\rm d}s^2=A(r) (\Xi  {\rm d} t-\omega _1 {\rm d} \phi _1-\omega _2 {\rm d} \phi _2)^2-\frac{{\rm d} r^2}{B(r)}\\
	-\frac{r^2 {\rm d} z^2}{l^2}-\frac{r^2}{l^4} \underset{i=1}{\overset{2}{\sum }}(\omega
	_i {\rm d} t-\Xi  l^2 {\rm d} \phi _i)^2.
	\end{multline}
	This metric \textit{has been directly derived from the vielbein}~\eqref{r4} and $\eta_{i j}=\text{diag}(1,\,-1,\,-1,\,-1,\,-1)$. It is different from the rotating metric suggested in Eq.~(14) of~\cite{3}, which is Eq.~\eqref{m14} of this work. The difference resides in the last term in Eq.~\eqref{m14} which, in this case ($N=5,\,n=2$), reduces to $-(r^2/l^2)(\omega_1{\rm d}\phi_2-\omega_2{\rm d}\phi_1)^2$.
	
	{Knowing the metric we evaluate $e_i{}^{\mu}$ by $e_i{}^{\mu}=\eta_{i j}g^{\mu \nu}e^j{}_{\nu}$.} Next, we evaluate $T$ upon substituting~\eqref{r4} and~\eqref{r4b} into~\eqref{o1} and the resulting expression is identical to~\eqref{T} taking $N=5$.
	
	Now, on substituting~\eqref{r4}, \eqref{r4b} and~\eqref{T} into the field equations~\eqref{o2} and using the static solution~\eqref{sol} we noticed that all the field equations are satisfied.
	
	{We have thus obtained a new rotating solution given by~\eqref{r4b}, which we rewrite for convenience
		\begin{multline}\label{diag}
		{\rm d}s^2=A(r) \Big(\Xi{\rm d}t-\sum_{i=1}^{2}\omega_i{\rm d}\phi_i\Big)^2-\frac{{\rm d} r^2}{B(r)}\\
		-\frac{r^2 {\rm d} z^2}{l^2}-\frac{r^2}{l^4} \underset{i=1}{\overset{2}{\sum }}(\omega
		_i {\rm d} t-\Xi  l^2 {\rm d} \phi _i)^2.
		\end{multline}	
		This is a solution to the field equations~\eqref{o2} with $e^i{}_{\mu}$ given by~\eqref{r4}, $\eta_{i j}=\text{diag}(1,\,-1,\,-1,\,-1,\,-1)$, $\Xi=\sqrt{1+\Sigma_{i=1}^2(\omega_i^2/l^2)}$, $A_\mu{\rm d}x^\mu=\Phi(r)(\Xi{\rm d}t-\Sigma_{i=1}^2\omega_i{\rm d}\phi_i)$, and the $r$-functions ($A,\,B,\,\Phi$) are given in~\eqref{sol}.}

	\subsubsection*{{Case b) $\eta_{i j}$ non-diagonal}}
	{The authors of Ref.~\cite{3} did not provide an expression for the Minkowskian metric $\eta_{i j}$ they used in their work. In our first version of this work we assumed $\eta_{i j}=\text{diag}(1,\,-1,\,-1,\,-1,\,-1)$ and we reached the conclusion that the metric~\eqref{m14} is not a solution to the field equations~\eqref{o2}. However, an anonymous referee claimed that a correct expression for $\eta_{i j}$ would be the matrix~(44) of Ref.~\cite{4}, which is also the matrix~(41) of Ref.~\cite{5}. The rightmost column and the bottom line of that matrix have a common element, which is $-1$, and the rest of the elements of the rightmost column and the bottom line are 0. In the case of five-dimensional spacetime with 2 angular coordinates ($N=5,\,n=2$), matrix~(44) of Ref.~\cite{4}, or matrix~(41) of Ref.~\cite{5}, takes the following form using the notation and signature of this work~\cite{fn}
		\begin{equation}\label{mat}
		\eta_{ij} = \begin{pmatrix}
		1 & 0 & 0 & 0 & 0 \\
		0 & -1 & 0 & 0 & 0 \\
		0 & 0 & -1-\frac{\omega _2^2}{l^2 \Xi ^2} & \frac{\omega _1 \omega _2}{l^2 \Xi ^2} & 0 \\
		0 & 0 & \frac{\omega _1 \omega _2}{l^2 \Xi ^2} & -1-\frac{\omega _1^2}{l^2 \Xi ^2} & 0 \\
		0 & 0 & 0 & 0 & -1
		\end{pmatrix}\,.
		\end{equation}
		With this $\eta_{ij}$ matrix and the expression of $e^i{}_{\mu}$ given in~\eqref{r4}, the formula $g_{\mu \nu} =
		\eta_{ij}e^{i}{_{\mu}}e^{j}{_{\nu}}$ yields the metric~\eqref{m14}. It is straightforward to show that the metric~\eqref{m14}, which we rewrite for convenience
		\begin{multline}\label{ndiag}
		{\rm d}s^2=A(r)\Big(\Xi{\rm d}t-\sum_{i=1}^{2}\omega_i{\rm d}\phi_i\Big)^2-\frac{{\rm d}r^2}{B(r)}-{{r^2\rm d}z^2\over l^2}\\
		\hspace{-1mm}-\frac{r^2}{l^4}\sum_{i=1}^{2}(\omega_i{\rm d}t-\Xi l^2{\rm d}\phi_i)^2-\frac{r^2}{l^2}(\omega_1{\rm d}\phi_2-\omega_2{\rm d}\phi_1)^2 ,
		\end{multline}
		is a solution to the field equations~\eqref{o2} with $e^i{}_{\mu}$ given by~\eqref{r4}, $\eta_{i j}$ given by~\eqref{mat}, $\Xi=\sqrt{1+\Sigma_{i=1}^2(\omega_i^2/l^2)}$, $A_\mu{\rm d}x^\mu=\Phi(r)(\Xi{\rm d}t-\Sigma_{i=1}^2\omega_i{\rm d}\phi_i)$, and the $r$-functions ($A,\,B,\,\Phi$) are given in~\eqref{sol}.}
	
	{It is also straightforward to show that $T$, upon substituting~\eqref{r4} and~\eqref{ndiag} into~\eqref{o1}, has the same expression as in~\eqref{T} taking $N=5$.}
	
	{In concluding, there are two cylindrically rotating solutions to the field equations~\eqref{o2}. The first solution, derived in this work~\eqref{diag}, is much simpler and is used with a diagonal Minkowskian metric $\eta_{i j}=\text{diag}(1,\,-1,\,-1,\,-1,\,-1)$. The second solution~\eqref{ndiag}, derived in Ref.~\cite{3} (with the global sign correction of its last term made in this work), includes \emph{extra terms}, $-(r^2/l^2)\sum_{i<j}^{n}(\omega_i{\rm d}\phi_j-\omega_j{\rm d}\phi_i)^2$, the number of which depends on the number $n$ of angular coordinates and is used with a non-diagonal Minkowskian metric $\eta_{ij}$~\eqref{mat}.}
	
	{It is not clear why the authors of Refs.~\cite{3,4,5} used a non-trivial, non-diagonal, Minkowskian metric~\eqref{mat} that they claim to be the {\textquoteleft Minkowskian metric in cylindrical coordinates\textquoteright}. This has nothing to do with cylindrical coordinates! (see~\cite{fn} for details). Moreover, such a non-diagonal Minkowskian metric has led to a more complicated rotating solution~\eqref{ndiag}. As a consequence, the rotating solutions derived in~\cite{4,5} have the same complicated structure as the one derived in~\cite{3} and they can be simplified on removing the extra terms $\mp (r^2/l^2)\sum_{i<j}^{n}(\omega_i{\rm d}\phi_j-\omega_j{\rm d}\phi_i)^2$ provided they are used with a diagonal Minkowskian metric $\eta_{i j}=\pm \text{diag}(1,\,-1,\,-1,\,-1,\,\cdots,-1)$.}  
	
	A point to emphasize is that when evaluating the metric from the formula $g_{\mu \nu} =
	\eta_{ij}e^{i}{_{\mu}}e^{j}{_{\nu}}$ one has to use $\eta_{i j}=\pm \text{diag}(1,\,-1,\,-1,\,-1,\,\cdots,-1)$ and not a non-diagonal expression. The tetrad defined in~\eqref{r4} forms a \emph{trivial pseudo-Cartesian system with metric} $\eta_{i j}=\text{diag}(1,\,-1,\,-1,\,-1,\,\cdots,-1)$. {Another anonymous referee has supported our claim.}
	
	{\section*{Non-diagonal solutions versus diagonal solutions}}
	{From now on, a non-diagonal Minkowskian metric will be denoted by $\bar{\eta}_{ij}$. Let $\bar{\eta}_{ij}$ and $\eta_{ij}$ be a non-diagonal and a diagonal Minkowskian metrics of dimension $N$, respectively. These two metrics may be related by a \emph{symmetric} matrix $R$ ($R_{ij}=R_{ji}$) such that $\bar{\eta}_{ij}={\eta}_{ik}R_{kj}$. For instance, $\bar{\eta}_{ij}$ given by~\eqref{mat} and $\eta_{i j}=\text{diag}(1,\,-1,\,-1,\,-1,\,-1)$ are related by $R_{ij}=\eta_{i k}\bar{\eta}_{kj}$:
		\begin{equation}\label{R}
		R_{ij} = \begin{pmatrix}
		1 & 0 & 0 & 0 & 0 \\
		0 & 1 & 0 & 0 & 0 \\
		0 & 0 & 1+\frac{\omega _2^2}{l^2 \Xi ^2} & -\frac{\omega _1 \omega _2}{l^2 \Xi ^2} & 0 \\
		0 & 0 & -\frac{\omega _1 \omega _2}{l^2 \Xi ^2} & 1+\frac{\omega _1^2}{l^2 \Xi ^2} & 0 \\
		0 & 0 & 0 & 0 & 1
		\end{pmatrix}\,.
		\end{equation}
		Let $\bar{g}_{\mu\nu}$ and $g_{\mu\nu}$ be the corresponding spacetime metrics, respectively.}
	
	{The purpose of this section is to show that if ($g_{\mu\nu},\,\eta_{ij}$) is a rotating solution then ($\bar{g}_{\mu\nu},\,\bar{\eta}_{ij}$) is a rotating solution too with similar geometrical properties. Using $\bar{g}_{\mu \nu} =
		\bar{\eta}_{ij}e^{i}{_{\mu}}e^{j}{_{\nu}}$ and the fact that $\bar{g}_{\mu \sigma}\bar{g}^{\sigma \nu}=\delta_\mu^\nu$ we obtain
		\begin{equation}\label{R1}
		\bar{g}^{\mu \nu}=\eta^{ik}R^{kj}e_i{}^{\mu}e_j{}^{\nu},
		\end{equation}
		where $\eta^{ik}$ and $R^{kj}$ are the inverse matrices of $\eta_{ik}$ and $R_{kj}$, respectively. Next, we evaluate $\bar{e}_i{}^{\mu}=\bar{\eta}_{i j}\bar{g}^{\mu \nu}e^j{}_{\nu}$. Using the expression~\eqref{R1} of $\bar{g}^{\mu \nu}$ and the fact that $R_{ij}$ is symmetric, we obtain
		\begin{equation}\label{R2}
		\bar{e}_i{}^{\mu}={e}_i{}^{\mu},
		\end{equation}
		which along with the relation $\bar{e}^{i}{_{\mu}}=e^{i}{_{\mu}}$ (true by definition since we are using the same tetrad but different Minkoskian metrics) imply that all the barred relevant entities entering the field equations~\eqref{o2} are equal to the non-barred entities. Hence, if the field equations are satisfied for the non-barred entities, they are automatically satisfied for the barred entities.}
	
	Our solution~\eqref{diag} includes four terms and the solution derived in Ref.~\cite{3}, Eq.~\eqref{ndiag}, includes the \emph{same four terms} plus the extra term $-\frac{r^2}{l^2}\sum_{i<j}^{n}(\omega_i{\rm d}\phi_j-\omega_j{\rm d}\phi_i)^2$, which in the case $N=5,\,n=2$ takes the form $-\frac{r^2}{l^2}(\omega_1{\rm d}\phi_2-\omega_2{\rm d}\phi_1)^2$. It is clear that these two solutions are manifestly different. Even if they share some similar geometrical and physical properties they are certainly different solutions because \emph{they cannot be related by a global coordinate transformation.}\\
	
	\section*{Concluding remarks}
	{We have thus shown that a trivial generalization of the transformation~\eqref{sub} to higher dimensional spacetimes is possible. By virtue of such a generalization we derived a simple cylindrically rotating solution of the form~\eqref{m14} with the last term $- (r^2/l^2)\sum_{i<j}^{n}(\omega_i{\rm d}\phi_j-\omega_j{\rm d}\phi_i)^2$ removed. This newly derived metric along with $A_\mu{\rm d}x^\mu=\Phi(r)(\Xi{\rm d}t-\Sigma_{i=1}^n\omega_i{\rm d}\phi_i)$ is a solution to the field equations~\eqref{o2} provided the Minkowskian metric is diagonal $\eta_{i j}= \text{diag}(1,\,-1,\,-1,\,-1,\,\cdots,-1)$ with the tetrad given by the expression~(12) of~\cite{3}. The $r$-functions ($A,\,B,\,\Phi$) are given in~\eqref{sol}.}
	
	{Another, non-trivial, generalization of~\eqref{sub} is also possible yielding a \emph{complicated} cylindrically rotating solution of the form~\eqref{m14}. This metric along with $A_\mu{\rm d}x^\mu=\Phi(r)(\Xi{\rm d}t-\Sigma_{i=1}^n\omega_i{\rm d}\phi_i)$ is a solution to the field equations~\eqref{o2} provided the Minkowskian metric is non-diagonal of the general form given in Eq.~(44) of Ref.~\cite{4} and Eq.~(41) of Ref.~\cite{5} with the tetrad given by the expression~(12) of~\cite{3}. The $r$-functions ($A,\,B,\,\Phi$) are given in~\eqref{sol}.}
	
	{We have also shown that if ($g_{\mu\nu},\,\eta_{ij}$) is a rotating solution with $\eta_{ij}$ being diagonal, then ($\bar{g}_{\mu\nu},\,\bar{\eta}_{ij}$) is another rotating solution with $\bar{\eta}_{ij}={\eta}_{ik}R_{kj}$ being non-diagonal and $R_{ij}$ is a symmetric matrix. These two rotating solutions have the same geometrical properties.}

	
	%
	


\end{document}